\documentclass[journal,final,twoside]{IEEEtran}

\usepackage{cite}
\usepackage{amsmath}
\usepackage{amssymb}
\usepackage{amsthm}
\usepackage{thmtools}
\usepackage{graphicx}
\usepackage{xcolor}
\usepackage{bbm}
\usepackage{hyperref}

\theoremstyle{definition}
\newtheorem{remark}{Remark}
\declaretheorem[style=remark,qed=$\blacktriangleleft$]{example}

\DeclareMathOperator{\coeff}{coeff}
\DeclareMathOperator{\argmin}{argmin}
\newcommand{\Mod}[1]{\ (\mathrm{mod}\ #1)}

\allowdisplaybreaks

\title{Performance-Complexity-Latency Trade-offs of Concatenated
RS-SDBCH Codes}
\author{Alvin Y. Sukmadji, \emph{Graduate Student Member, IEEE}
and Frank R. Kschischang, \emph{Fellow, IEEE}\thanks{Submitted on
October 28, 2024, revised on January 10, 2025,
accepted on January 14, 2025. 
The authors are with the Edward S.\ Rogers Sr.\
Department of Electrical \& Computer Engineering, University of
Toronto, Canada.  Emails:
alvin.sukmadji@mail.utoronto.ca,
frank@ece.utoronto.ca.
Parts of this work have been accepted for a presentation
at the 2025 Optical Fiber Commun. Conf. (OFC) \cite{sukmadji-ofc25}.}}

\begin{document}
\maketitle
\begin{abstract}
Concatenated bit-interleaved and multilevel coded modulation with
outer Reed--Solomon codes, inner Chase-algorithm-based
soft-decision-decoded Bose--Ray-Chaudhuri--Hocquenghem codes, and
four-level pulse amplitude modulation is considered.  A
semi-analytical formula is derived for estimating the decoded frame
error rate (FER) at the output of the additive white Gaussian noise
channel, obviating the need for time-consuming Monte Carlo
simulations.  The formula is used to search a large space of codes
(including the KP4 code) to find those achieving good trade-offs
among performance (measured by the gap to the constrained Shannon limit at
$10^{-13}$ FER), complexity (measured by the number of elementary
decoder operations), and latency (measured by overall block
length).
\end{abstract}


\section{Introduction}
\label{sec:intro}
\IEEEPARstart{T}{his} paper examines the
performance-complexity-latency trade-offs achieved by coded
modulation schemes based on four-level pulse amplitude modulation
(PAM4) using concatenated outer Reed--Solomon (RS) outer codes and
soft-decision (SD) Bose--Ray-Chaudhuri--Hocquenghem (BCH) inner
codes.  Bit-interleaved coded modulation (BICM) and multilevel
coded (MLC) modulation are both considered.  The use of RS-SDBCH
codes is motivated by potential applications in short-reach
high-throughput optical communication systems, such as data center
interconnects, where low-complexity low-latency forward error
correction (FEC) is essential.

%With the ever increasing Internet traffic, the demand for
%higher throughput communication systems also goes up,
%and it is sensible that a higher-order modulation scheme
%is used in many communication systems.
The two-level non-return-to-zero (NRZ) modulation scheme has in
recent years been supplanted by PAM4 in many optical transmission
systems due to its higher bit rate and spectral
efficiency~\cite{chang,troncoso,buchali,vanveen,agrell2024}, and
indeed PAM4 is the recommended transmission format in various
Ethernet standards at transmission rates of up to 400 Gb/s per
lane \cite{ieee8023df,nagarajan,willner}.  In a short-haul
applications, PAM4 demonstrates lower propagation penalties than
NRZ with only a modest increase in the optical power
\cite{szczerba}, which makes PAM4 modulation scheme attractive for
data center interconnects.

In transmission systems involving higher-order modulation, coded
modulation schemes are often employed. Two generic coded
modulation architectures that we will consider in this paper are
bit-interleaved coded modulation (BICM) \cite{zehavi,caire} and
multilevel coding (MLC) \cite{wachsmann,ungerboeck,imai}.  In
BICM, all FEC bits are interleaved prior to being mapped (usually
according to a Gray code) to transmitted constellation points.
Though not generally capacity-achieving \cite{caire}, BICM is
often considered the pragmatic choice in many
applications~\cite{smith}.  MLC, on the other hand, can in
principle achieve capacity by separately coding each of the
bit-level channels induced by a specific constellation labelling,
using multi-stage decoding (MSD) in keeping with the chain rule of
mutual information to avoid information loss \cite{wachsmann}.
The need to implement multiple decoders, and the latency and
potential error-propagation induced by MSD are often seen as major
drawbacks of MLC schemes \cite{wachsmann,ungerboeck}.

A PAM4 constellation, however, has only two bit levels, and it
becomes possible to consider MLC schemes with a natural binary
labelling in which only the least significant bit (LSB) level is
protected by an inner FEC while the most significant bit (MSB)
level is left uncoded (though still conditionally demapped,
thereby benefitting from a 6~dB effective signal-to-noise-ratio
enhancement)~\cite{barakatain-higherordermodulation,mehmood,yoshida}.
Errors in the MSB channel caused by channel noise or resulting
from a propagation of errors remaining in the LSB channel after
inner decoding are then passed to the outer code.  Since only a
fraction of transmitted bits must be decoded by the inner code,
such an MLC scheme may offer complexity benefits compared with
BICM.

In our previous work \cite{sukmadji2024}, we have studied
performance-complexity-latency trade-offs of concatenated RS outer
and hard-decision BCH inner codes over the binary symmetric
channel.  Here we expand upon this previous work by considering
the use of soft-decision inner BCH codes using PAM4 with BICM and
MLC coded modulation scheme over the additive white Gaussian noise
(AWGN) channel.  We consider the Chase (or Chase-II) decoder
\cite{chase} to decode the inner BCH codes.  Concatenated RS-SDBCH
codes have been used or considered in many transmission schemes
\cite{lentner23,yang,wang2023}. These papers consider schemes
using the so-called KP4 code, a 15-error-correcting
RS(544, 514) code, as the outer
code, as this code is embedded in the electrical interface in many
Ethernet and data center
interconnects~\cite{ieee8023,bhoja,chagnon}.  While we pay close
attention to the performance-complexity-latency tradeoffs
achievable when the KP4 code is chosen as the outer code, in this
paper we consider also (many) other possible choices of outer RS
code.

\begin{figure*}[t]
\centering
\includegraphics{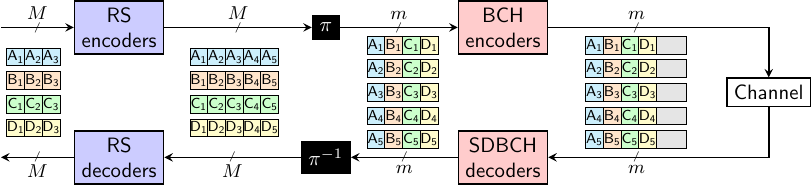}
\caption{Example of an RS-SDBCH concatenation with $M=4$ RS
codewords and $m=5$ BCH codewords with the ``card-dealing''
interleaving scheme.  Each square represents one ($B$-bit) RS
symbol and the number in each square represents the RS codeword
index from which that symbol originates.  Here, $K=3$, $N=5$,
$k=4B $.}
\label{fig:system}
\end{figure*}

The remainder of this paper is organized as follows.  In
Sec.~\ref{sec:model}, we describe the class of RS-SDBCH
concatenated coding schemes considered in this paper and in
Sec.~\ref{sec:bicm-mlc} we describe how such schemes can be used
in conjunction with PAM4 modulation according to the BICM and MLC
architectures.  In Sec.~\ref{sec:estimating-fer} we provide a
semi-analytical formula for estimating the post-FEC frame error rate
(FER) of such schemes.  We used this formula to search a large
space of codes to find parameter-combinations achieving good
performance-complexity-latency trade-offs.  The results of this
search are given in Sec.~\ref{sec:pcl-tradeoffs} and also posted
online~\cite{sukmadji-pareto}.

Throughout this paper, for any positive integer $n$, we let
$[n]=\{0,1,\ldots,n-1\}$, with $[0]=\varnothing$.  The finite
field with $q$ elements is denoted as $\mathbb{F}_q$.  The
notation $\mathcal{C}(n,k,t)$ refers to a linear code
$\mathcal{C}$ with block length $n$, dimension $k$, and
error-correcting radius $t$.  Finally, ``$\oplus$'' denotes the
bit-wise XOR operator.

\section{System Model}
\label{sec:model}

Similar to the system in \cite[Fig.~1]{sukmadji2024}, we consider
a concatenated coding system consisting of $M$ outer RS($N,K,T$)
codes with $B$ bits per symbol interleaved with $m$ inner
BCH($n,k,t$) codes, shortened from a BCH code of length $2^b-1$,
or from $2^b$ in the case of extended BCH (eBCH), for some
positive integer $b$.  We assume that both inner and outer codes
are encoded systematically. That is, the first $K$ symbols of an
RS codeword are information symbols and similarly, the first $k$
bits of a BCH codeword are information bits.

We assume that the interleaving, i.e., the mapping of symbols,
between inner and outer codes occurs in a symbol-wise manner. That
is, all bits of a given RS symbol are interleaved into the
information positions of a single inner BCH code. The interleaver
induces an $M\times m$ \emph{adjacency matrix} $\mathbf{L}$, with
entries in the $i$th row and $j$th column, $L_{i,j}$, denoting the
number of symbols interleaved from the $i$th outer RS codeword to
the $j$th inner BCH codeword for all $i\in[M]$ and $j\in[m]$.
Thus, the balance condition
\begin{align}
\sum_{j=0}^{m-1}L_{i,j}=N
\label{eqn:balance-cond1}
\end{align}
must be satisfied.

To ensure that an (approximately) equal number of RS symbols are
interleaved into each of the inner codewords, we use a
``\emph{card-dealing}'' interleaving scheme, in which the $j$th
symbol of the $i$th RS codeword is interleaved into the
information position of the $((Ni+j)\Mod{m})$th inner codeword.
Thus, each entry of the adjacency matrix is either
$L_{i,j}=\lfloor N/m\rfloor$ or $\lceil N/m\rceil$ for all
$i\in[M]$, $j\in[m]$. An example of the concatenated system with a
card-dealing interleaver is shown in Fig.~\ref{fig:system}.

The inner codewords are modulated according to the PAM4 modulation
scheme with BICM and MLC schemes before they are transmitted
through the channel. The modulation scheme is described in detail
in Sec.~\ref{sec:bicm-mlc}.

For simplicity, we assume that all $L_{i,j}$ RS symbols
interleaved from the $i$th RS codeword to the $j$th BCH codeword
are placed adjacent to each other in the corresponding BCH
codeword, forming a \emph{strip} of length $L_{i,j}$ RS symbols.

% \begin{remark}
% In practice, the outer RS encoder and decoder are typically placed
% in the host chip, while the inner BCH encoder and decoder are
% connected by a chip-to-module electrical interface (C2M). The C2M
% may introduce more bit errors, on top of that from the channel,
% with a bit error rate in the
% order of $10^{-5}$ \cite{lusted,li}. For simplicity,
% we do not consider the bit errors that arise from the C2M in this
% paper.
% \label{rem:c2m}
% \end{remark}

\section{BICM and MLC for the Inner Codes}
\label{sec:bicm-mlc}

For the remainder of the paper, we assume that $B$ is even and
each RS symbol contains $B/2$ PAM4 symbols.  For each pair of bits
in a PAM4 symbol, we call the first and second bits to be the most
significant bit (MSB) and least significant bit (LSB),
respectively.

\subsection{Bit-Interleaved Coded Modulation}
\label{sec:bicm}
In BICM, we assume that $B$ divides $k$ and in addition to
\eqref{eqn:balance-cond1}, another balance condition
\begin{align}
B\sum_{i=0}^{M-1}L_{i,j}=k
\label{eq:balance-cond-bicm}
\end{align}
must also be satisfied.
A Gray-labeled PAM4 constellation is used with a mapping between a
pair of bits to their respective constellation symbol of
$00\mapsto -3$, $01\mapsto -1$, $11\mapsto +1$, $10\mapsto +3$.
The $k$ BCH information bits (which are composed of $k/B$ RS
symbols) are encoded into $k/2$ PAM4 symbols, and the $n-k$ BCH
parity bits are encoded into
$\left\lceil\frac{n-k}{2}\right\rceil$ PAM4 symbols.  If $n-k$ is
odd, then one of the parity bits will be paired up with a frozen
zero bit, with the parity bit as the MSB and the zero bit as the
LSB.

There are $m$ BCH codewords being transmitted over the channel, or
equivalently, $m\cdot\left\lceil \frac{n}{2}\right\rceil$ PAM4
symbols.  The overall inner code rate for the BICM scheme is
$k/n$.

\begin{figure}[t]
\centering
\includegraphics{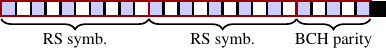}
\caption{Example of a BICM codeword with $n=25$, $k=20$ and
$B=10$. The shaded and unshaded squares represent the MSBs and
LSBs respectively. The rightmost black square is the ``zero-pad''
LSB paired with one of the parity MSBs.}
\label{fig:bicm-example}
\end{figure}

\begin{example}
Fig.~\ref{fig:bicm-example} shows an example of a BCH codeword
with $n=25$, $k=20$, and $B=10$.  There are $k/B=2$ RS symbols
with the first 10 bits assigned to the first symbol and the next
10 bits for the second symbol. Each RS symbol contains $10/2=5$
PAM4 symbols. Finally, since there are five parity bits, we have
three parity MSBs, while the remaining two are LSBs.  One of the
parity MSBs is paired with a ``zero-pad'' LSB.
\end{example}

\subsection{Multilevel Coding}
In MLC, we assume that $B/2$ divides $k$ and in addition to
\eqref{eqn:balance-cond1}, the balance condition
\begin{align}
\dfrac{B}{2}\sum_{i=0}^{M-1}L_{i,j}=k
\label{eq:balance-cond-mlc}
\end{align}
must also be satisfied.  Each bit in the inner codeword is paired
with one of the $k$ bits not protected by the inner code.  The $k$
BCH information bits and the $k$ bits not protected by the inner
code are made of the LSBs and MSBs, respectively, of some $2k/B$
RS symbols.

For the BCH information bits and bits not protected by the inner
code, we use a natural-labeled PAM4 constellation with mapping of
$00\mapsto -3$, $01\mapsto -1$, $10\mapsto +1$, $11\mapsto +3$.
The $n-k$ BCH parity bits, on the other hand, are modulated using
Gray-labeled PAM4 with BICM, similar to that in
Sec.~\ref{sec:bicm}.  If $n-k$ is odd, then one of the parity bits
is paired up with a frozen zero bit.

\begin{figure}[t]
\centering
\includegraphics{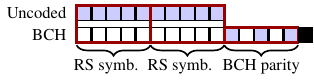}
\caption{Example of an MLC codeword with $n=15$, $k=10$ and
$B=10$. The shaded and unshaded squares represent the MSBs and
LSBs respectively. The rightmost BCH black square is the
``zero-pad'' LSB for the rightmost MSB.}
\label{fig:mlc-example}
\end{figure}

\begin{example}
Fig.~\ref{fig:mlc-example} shows an example of an inner BCH
codeword with $n=15$, $k=10$, and $B=10$ and 10 bits not protected
by the inner code. There are $2\cdot 10/10=2$ RS symbols that are
formed by the unprotected MSBs and BCH information LSBs.  The
first five unprotected bits and five BCH information bits form the
first RS symbol, while the next five unprotected bits and five BCH
information bits form the other RS symbol.  The five BCH parity
bits comprises three MSBs and two LSBs, plus one zero-pad to be
paired up with one of the parity MSBs.
\end{example}

\begin{figure}[t]
\centering
\includegraphics{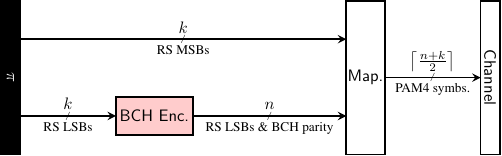}\\[2mm]
\includegraphics{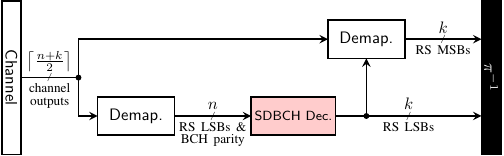}
\caption{Encoding (top) and decoding (bottom) schemes for MLC with PAM4.}
\label{fig:mlc-scheme}
\end{figure}

At the receiver, the MSBs and LSBs of the interleaved RS symbols
are recovered via MSD. First, the $n$ bits of the received BCH
words are obtained by demodulating and demapping the channel
outputs. Those bits are then decoded, and the decoded $k$
information bits (which are composed of the LSBs of the RS words),
are used to conditionally demodulate and demap the $k$ MSBs of the
RS symbols. The encoding and decoding schemes of MLC is summarized
in Fig.~\ref{fig:mlc-scheme}.

There are two main reasons why the MSBs and LSBs are assigned
as described above:
\begin{enumerate}
\item For the natural-labeled portion (RS MSBs and BCH information bits),
when the BCH decoder produces an LSB error,
roughly 1/3 of the time the corresponding MSB will then also
be demapped in error.  To minimize RS-symbol errors, it
is desirable to cluster these
correlated bit errors into the same RS symbol.
\item BCH parity bits are not passed to the RS decoder, thus
any errors occurring in these positions do not propagate to the outer decoder.
\end{enumerate}

In this scheme a total of $m$ BCH codewords plus $mk$ unprotected
bits are transmitted over the channel, or equivalently,
$m\cdot\left\lceil\frac{n+k}{2}\right\rceil$ PAM4 symbols. The
overall inner code rate for the MLC scheme is $\frac{2k}{k+n}$.

\section{Estimating the Frame Error Rate (FER)}
\label{sec:estimating-fer}

Throughout this section, we will mainly describe the FER
estimation for BICM.  The method for estimating the FER
in the case of MLC is very similar to that of BICM and the key
differences in each step will be pointed out.

\subsection{Overview}
\label{sec:dec-architecture}

\begin{figure}[t]
\centering
\includegraphics{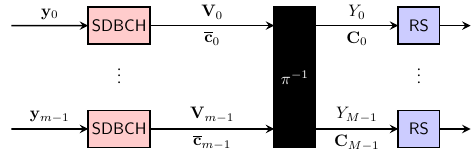}
\caption{Decoding scheme of the concatenated system. In the case
of MLC, the SDBCH decoder includes the conditional demodulator and
demapper as shown in Fig.~\ref{fig:mlc-scheme}.}
\label{fig:decoder}
\end{figure}

We consider the decoding scheme shown in Fig.~\ref{fig:decoder}.
The channel outputs $m$ vectors
$\mathbf{y}_0,\ldots,\mathbf{y}_{m-1}\in\mathbb{R}^{\left\lceil
n/2\right\rceil}$, which are demapped and decoded using $m$
parallel SDBCH decoders.  For all $j\in[m]$, the SDBCH decoder
outputs the BCH information bits
\[
\overline{\mathbf{c}}_j=\left(\overline{\mathbf{c}}_{0,j},
          \ldots,\overline{\mathbf{c}}_{M-1,j}\right)\in\mathbb{F}_2^k,
\]
where for all $i\in[M]$,
$\overline{\mathbf{c}}_{i,j}\in\mathbb{F}_2^{BL_{i,j}}$. In other
words, $\overline{\mathbf{c}}_{i,j}$ is the strip of $L_{i,j}$ RS
symbols from the $i$th RS codeword that is interleaved to the
$j$th BCH codeword.  The number of RS-symbol errors within the
strips of $\overline{\mathbf{c}}_{j}$ is represented by a vector
random variable
\[
\mathbf{V}_j=\left(V_{0,j},\ldots,V_{M-1,j}\right),
\]
where $V_{i,j}\in[L_{i,j}+1]$ denotes the number of RS-symbol
errors in $\overline{\mathbf{c}}_{i,j}$.

The vector
$\overline{\mathbf{c}}_{0},\ldots,\overline{\mathbf{c}}_{m-1}$ is
then deinterleaved to form the received RS words
$\mathbf{C}_0,\ldots,\mathbf{C}_{M-1}$, with $\mathbf{C}_i$ having
$Y_i=V_{i,0}+\cdots+V_{i,m-1}$ RS-symbol errors for all $i\in[M]$.

\begin{remark}
In the MLC case, for all $j\in[m]$, we have
$\mathbf{y}_0,\ldots,\mathbf{y}_{m-1}\in\mathbb{R}^{\left\lceil
(n+k)/2\right\rceil}$ and $\overline{\mathbf{c}}_j$ is composed of
$2k$ bits: $k$ decoded BCH information bits and $k$
conditionally-demapped unprotected bits.
\label{rem:dec-architecture}
\end{remark}

\subsection{Soft Decoding of the Inner Codes}

We use Chase (or Chase-II) decoding \cite[Alg.~2]{chase} with $J$
test bits to decode inner BCH codes.  To decode the $j$th channel
output, $j\in[m]$, the algorithm proceeds as follows.  From the
channel output $\mathbf{y}_j$, the log-likelihood ratio (LLR) is
computed for each of the $n$ bits
$\boldsymbol{\ell}=(\ell_{0},\ldots,\ell_{n-1})$.  Next, the
hard-decision word $\mathbf{x}=\left(x_0,\ldots,x_{n-1}\right)$ is
extracted from the sign of the LLRs.  The decoder then determines the
set of positions of the $J$ least reliable bit positions
$\mathcal{I}\subseteq [n]$, $|\mathcal{I}|=J$, i.e., $J$ positions
where $|\ell_q| \leq |\ell_{q'}|$ for all $q\in \mathcal{I}$ and
$q'\in [n]\setminus \mathcal{I}$.

Next, $2^J$ test patterns
$\mathbf{e}_0,\ldots,\mathbf{e}_{2^J-1}\in\mathbb{F}_2^n$ are
generated with all $2^J$ possible combinations of zeros and ones
for bits in positions in $\mathcal{I}$ and zeros everywhere else.
The BCH decoder then performs hard-decision decoding on each of
the vectors
$\mathbf{x}\oplus\mathbf{e}_0,\ldots,\mathbf{x}\oplus\mathbf{e}_{2^J-1}$.
For all $q\in\left[2^J\right]$, if decoding
$\mathbf{x}\oplus\mathbf{e}_q$ does not end in a decoding failure,
then we let
$\overline{\mathbf{x}}_{q}=\left(\overline{x}_{q,0},\ldots,\overline{x}_{q,n-1}\right)$
be the output codeword. The corresponding
\emph{analog weight} is then computed as
\[
W_q=\sum_{i=0}^{n-1}(x_{q,i}\oplus\overline{x}_{q,i})\cdot|\ell_i|.
\]
On the other hand, if the decoding fails, the analog weight is
set to $W_q=+\infty$.

If $W_q=+\infty$ for all $q\in\left[2^J\right]$, then all decoding attempts
have failed and the decoder will output
$\overline{\mathbf{c}}_j=\left(x_{0},\ldots,x_{k-1}\right)$,
the information bits of the initial hard-decision word.
Otherwise, the Chase decoder determines the index
of the lowest analog weight $q^*$, i.e., $q^*=\argmin_{q}W_q$, and
outputs
\[
\overline{\mathbf{c}}_j=\left(\overline{x}_{q^*,0},\ldots,\overline{x}_{q^*,k-1}\right),
\]
the information bits of the codeword $\overline{\mathbf{x}}_{q^*}$.

\begin{remark}
In the MLC case, Chase decoding is performed only on the $k$ BCH
information bits (i.e., LSBs of the RS symbols) and the $n-k$ BCH
parity bits.  Also, as pointed out in
Remark~\ref{rem:dec-architecture}, $\overline{\mathbf{c}}_j$
comprises $2k$ bits: $k$ information bits of the decoded BCH
codeword (or initial hard-decision word
in the event of no successful decoding attempts)
and $k$ conditionally-demapped unprotected bits.
\end{remark}

\subsection{RS-Symbol-Error Weight Distribution}
\label{sec:symb-weight-dist}

Consider a strip of $L$ RS-symbols (or $BL$ bits) in the $k$
information bits of a BCH (code)word.  Suppose that the BCH
information bits contain $u\in[k+1]$ bit errors that induce
$v\in[L+1]$ RS-symbol errors in the strip.  We have shown in
\cite[eq.~(4)]{sukmadji2024} that the number of possible $u$
bit-error patterns in the information bits that induce $v$
RS-symbol errors in that strip of $L$ RS symbols is the
coefficient of $x^uy^v$ of the bivariate generating function
\begin{align}
W_{B,k,L}(x,y)=\left(1+\left((1+x)^{B}-1\right)y\right)^L(1+x)^{k-BL}.
\label{eqn:genfunc}
\end{align}
The indeterminate $x$ of \eqref{eqn:genfunc} keeps track of the
bit errors within the BCH information bits, while the
indeterminate $y$ keeps track of the RS-symbol errors in the
strip.

We generalize the definition of the generating function
\eqref{eqn:genfunc} so that the indeterminate $x$ keeps track of
the two-bit PAM4-symbol errors within the information bits.  In
the BICM scheme described in Sec.~\ref{sec:bicm}, each symbol may
contain up to $B/2$ PAM4 symbol errors and the information bits
contain $k/2$ PAM4 symbols.  The corresponding generating function
keeping track of the PAM4-symbol errors and the induced RS-symbol
errors in a strip made of $L$ RS symbols is therefore
$W_{B/2,k/2,L}(x,y)$.
% For the BICM scheme, the information bits contain $k/2$ PAM4 symbols, while for the MLC scheme,
% the uncoded and BCH information bits combined contain $k$ PAM4 symbols.
% Thus, the corresponding generating functions for BICM and MLC are $W_{B/2,k/2,L}(x,y)$
% and $W_{B/2,k,L}(x,y)$, respectively.

\begin{remark}
In this generalization, $x$ keeps track whether a PAM4 symbol is
in error or not. However, $x$ does not keep track of the actual
bit-error patterns in the PAM4 symbol in the event of a
PAM4-symbol error (i.e., $01$, $10$, or $11$).
\end{remark}

Let $U_j$ denote the number of PAM4-symbol-error weight in
$\overline{\mathbf{c}}_j$.  Assuming that any PAM4-symbol-error
patterns of weight $u$ is equally likely to occur in
$\overline{\mathbf{c}}_j$, we can determine the conditional
RS-symbol-error weight distribution of strip
$\overline{\mathbf{c}}_{i,j}$ as
\begin{align}
\Pr(V_{i,j}=v\mid U_j=u)=\frac{\coeff_{x^uy^v}(W_{B/2,k/2,L_{i,j}}(x,y))}{\binom{k/2}{u}}
\label{eqn:cond-symb-weight}
\end{align}
for all $u\in\left[\frac{k}{2}+1\right]$ and
$v\in\left[L_{i,j}+1\right]$.
Then, using the law of total probability, we obtain the
RS-symbol-error weight distribution of the strip as
\begin{align}
\displaystyle\Pr(V_{i,j}=v)=\sum_{u=0}^{k/2}\Pr(V_{i,j}=v\mid U_j=u)\Pr(U_j=u).
\label{eqn:symb-err-dist}
\end{align}
for all $u\in\left[\frac{k}{2}+1\right]$ and $v\in\left[L_{i,j}+1\right]$.

We note that the distribution of $U_j$ is difficult to estimate
analytically.  We estimate this distribution using Monte Carlo
simulations over varying inner code and channel parameters.  Hence
we have referred to our approach as being ``semi-analytical.''

\begin{remark}
In the MLC case, the corresponding generating function keeping
track of the PAM4-symbol errors and the induced RS-symbol errors
in a strip of $L$ RS symbols is $W_{B/2,k,L}(x,y)$ since
$\overline{\mathbf{c}}_j$ contains $2k$ bits, and therefore, $k$
PAM4 symbols.  Thus, \eqref{eqn:cond-symb-weight} and
\eqref{eqn:symb-err-dist} become
\begin{align}
\Pr(V_{i,j}=v\mid U_j=u)=\frac{\coeff_{x^uy^v}(W_{B/2,k,L_{i,j}}(x,y))}{\binom{k}{u}}
\label{eqn:cond-symb-weight2}
\end{align}
and
\begin{align}
\displaystyle\Pr(V_{i,j}=v)=\sum_{u=0}^{k}\Pr(V_{i,j}=v\mid U_j=u)\Pr(U_j=u),
\label{eqn:symb-err-dist2}
\end{align}
respectively, for all $u\in\left[k+1\right]$ and $v\in\left[L_{i,j}+1\right]$.
\end{remark}

\subsection{Estimating the FER}

After obtaining the RS-symbol-error weight distribution for each
strip, the method of estimating the FER is identical to that in
\cite[eqs.~(7)--(9)]{sukmadji2024}.  Recall from
Sec.~\ref{sec:dec-architecture} that the total number of RS-symbol
errors of the $i$th RS word is
$Y_i=V_{i,0}+V_{i,1}+\cdots+V_{i,m-1}$.  For a fixed $i$, the
random variables $V_{i,0},\ldots.V_{i,m-1}$ are independent since
they come from independent inner codes.  Hence, the distribution
of $Y_i$ is a convolution of the $V_{i,j}$'s, $j\in[m]$.

A \emph{frame error} is defined as an event where at least one of
the received RS words contain more than $T$ RS-symbol errors. The
frame error rate (FER) is therefore given as
\[
\Pr(\text{frame error}) = \Pr\left(Y_0>T \vee \cdots \vee Y_{M-1}>T\right).
\]
The probability above is typically difficult to estimate due to
the correlation between the symbols across different RS words.
However, we can estimate the upper bound using a simple union
bound
\begin{equation}
\Pr(\text{frame error}) \leq \sum_{i=0}^{M-1}\Pr\left(Y_i>T\right).
\label{eq:unionbound}
\end{equation}

\begin{figure}[t]
\centering
\includegraphics{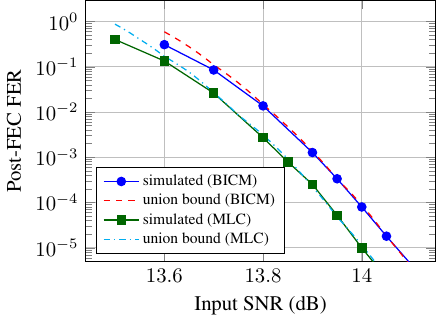}
\caption{FER versus input SNR for concatenated
$10\times\,\text{RS}(544,514,15)$ (i.e., KP4) outer codes and
$544\times\,\text{eBCH}(115,100,2)$ inner codes for BICM, and
$544\times\,\text{eBCH}(65,50,2)$ inner codes for MLC ($J=2$ in
both cases) over the AWGN channel.}
\label{fig:fer-bicm}
\end{figure}

Fig.~\ref{fig:fer-bicm} show the simulation results and union
bound estimates of the post-FEC FER based on the steps outlined in
this section for BICM and MLC, respectively.  In both cases we
observe gaps between the FERs obtained from simulation and
estimation at low input signal-to-noise ratio (SNR). However,
those gaps are closed when the SNR is sufficiently high.

\begin{remark}
While we have derived an FER estimate, the formula can be easily
extended to estimate the post-FEC bit error rate (BER).  In the
event of a frame error for our input SNR regime of interest, it is
very likely that only one (out of $M$) RS word contains more than
$T$ RS-symbol errors. Furthermore, the most likely RS-symbol-error
weight in a frame error is that of lowest possible weight $T+1$.
We can also assume that the positions of the bits in error are
spaced apart such that for every RS-symbol that is in error, only one
PAM4 symbol is in error. In such an error pattern there are, on average,
$(1 + \gamma)(T+1)$ bit errors out of $MNB$ bits, giving the
BER estimate
\[
\frac{(1 + \gamma)(T+1)}{MNB} \cdot \text{FER},
\]
where $\gamma = 0$ for BICM and $\gamma \approx 0.3$ for MLC.
\end{remark}

\section{Performance-Complexity-Latency Trade-offs}
\label{sec:pcl-tradeoffs}

\subsection{Figures of Merit}
\label{sec:metric}
\subsubsection{Performance}
\label{sec:performance}
Performance is defined to be the gap---measured in dB---to the PAM4 constrained
Shannon limit (CSL) at $10^{-13}$ post-FEC FER
over the additive white Gaussian noise channel.

\subsubsection{Complexity}
\label{sec:complexity}
Complexity is defined to be the worst-case number of elementary
operations performed by the inner and outer decoders normalized by
the number of decoded information bits. Here, elementary decoding
operations constitute additions and multiplications (over real
numbers and integers), comparisons, and table lookups.

The complexity formula for the concatenated coding scheme is
\[
\frac{M\kappa_{\text{out}}+m\kappa_{\text{in}}}{MKB},
\]
where $\kappa_{\text{out}}$ denotes the number of elementary
operations needed to decode an outer RS word and
$\kappa_{\text{in}}$ denotes the number of elementary operations
needed to decode a BCH word using a Chase decoder with $J$ test
bits.  Similar to the assumptions made in
\cite[Sec.~V-A2]{sukmadji2024}, for the RS and hard-decision BCH
decoding, we assume a simplified decoding procedure for $T$ or
$t=1,2,3,4$, and a general-purpose inversionless Berlekamp--Massey
algorithm and Chien search for $T$ or $t\geq 5$.

The formula of $\kappa_{\text{out}}$ is identical to that of
$\kappa_{\text{RS}}$ defined in \cite[Sec.~V-A2]{sukmadji2024}.
The complexity of decoding an outer RS($N,K,T$) code is
\begin{align*}
\kappa_{\text{out}}=&6K(N-K)+\text{KE}_{\text{RS}}(T)+\text{RF}_{\text{RS}}(T)\\
 &+6T\left(2T+\left\lceil\frac{T}{2}\right\rceil\right)-4T,
\end{align*}
where the values of $\text{KE}_{\text{RS}}(T)$ and
$\text{RF}_{\text{RS}}(T)$ are shown in
Table~\ref{tab:rs-complexity}.

\begin{table}[t]
\centering
\caption{Values of $\text{KE}_{\text{RS}}(T)$ and $\text{RF}_{\text{RS}}(T)$}
\begin{tabular}{|c|c|c|c|c|c|}\hline
$T$ & $1$ & $2$ & $3$ & $4$ & $\geq 5$\\\hline
$\text{KE}_{\text{RS}}(T)$ & 9 & 54 & 159 & 336 & $2T(24T+8)$\\\hline
$\text{RF}_{\text{RS}}(T)$ & 0 & 10 & 37 & 98 & $6NT$\\\hline
\end{tabular}
\label{tab:rs-complexity}
\end{table}

Decoding an inner (e)BCH($n,k,t$) word using a Chase decoder with
$J$ test bits requires the following operations (with the
corresponding number of elementary operations):
\begin{enumerate}
 \item LLR and hard-decision word computation: $2n$,
 \item finding the indices of the $J$ least reliable bits:
\[
(n-J)(J+5)+\frac{J(J-1)}{2}+3J,
\]
 \item test pattern generation: $2^J-1$,
 \item initial syndrome generation: $nt$
 \item test syndrome generation: $t\left(2^J-1\right)$,
 \item key equation solver: $0,2^J \cdot 11,2^J \cdot 23$ for
$t=1,2,3$, respectively,
 \item polynomial root finding: $0,2^J \cdot 10,2^J\cdot 37$ for
$t=1,2,3$, respectively,
 \item analog weight computation:
\[
	\max\left\{\sum_{s=0}^{\left\lfloor J/2\right\rfloor} \binom{J}{2s}(2s),\sum_{s=0}^{\left\lfloor J/2\right\rfloor} \binom{J}{2s+1}(2s+1)\right\}
\]
	for $t=1$, or
\[
	\sum_{s=0}^J \binom{J}{s}(t+s-1)
\]
	for $t=2,3$,
	\item finding the lowest analog weight: $2^{J-1}-1$ for $t=1$, or $2^J-1$ for $t=2,3$,
	\item bit correction: $J+t$, and
	\item demapping the MSBs (for MLC only): $2k$.
\end{enumerate}
Thus, $\kappa_{\text{in}}$ is the sum of the total number of
elementary operations for the first 10 items listed above in the
case of BICM, or all 11 items in the case of MLC.
Appendix~\ref{app:chase-complexity} describes each of the 11
operations listed above in greater detail.

We also consider the case where the inner codes are the single
parity check SPC($n,n-1$) codes with Wagner decoding
\cite{silverman}.  In this case, there are a total of
$\kappa_{\text{{in}}}=4n$ decoding operations in total
\cite[eq.~(19)]{qu2024}.  Appendix~\ref{app:wagner} describes
Wagner decoding in greater detail.

\subsubsection{Latency}
Following the reasoning given in \cite[Sec.~V-A3]{sukmadji2024},
we define latency to be the amount of time between when a
particular information bit arrives at the input of the decoder and
when that bit leaves the decoder. We assume that the decoding time
is equal to the time it takes for the incoming bits to fill in the
buffer at the receiver (so that the decoder never goes idle).  The
required time to fill in the buffer is proportional to the block
length $mn$ for BICM and $m(n+k)$ for MLC, which we take as the
metric for latency.

\subsection{Search Criteria}
\label{sec:criteria}
We perform a code search with the following criteria:
\begin{itemize}
\item RS outer codes with $T=1,\ldots,20$ and $B=10$.
\item Shortened $t=1,2,3$ eBCH inner codes with rates $\leq$ 0.99,
shortened from code of length $2^b$, $b=5,\ldots,11$ decoded using
Chase decoder with $J=1,\ldots,6$.
\item SPC codes ($t=0$, $b=$ ---, $J=$ ---) with rates $\leq$ 0.99
decoded using Wagner decoder.
\item Card-dealing interleaver between inner and outer codes.
\item Overall code rates of 0.75, 0.76,\,\ldots, 0.93 ($\pm$ 0.005).
\item Maximum latency of 20, 60, and 200 kilobits.
\item The balance conditions \eqref{eqn:balance-cond1} and
\eqref{eq:balance-cond-bicm} must be satisfied for BICM, or
\eqref{eqn:balance-cond1} and \eqref{eq:balance-cond-mlc} for MLC.
\item AWGN channel.
\end{itemize} 

As pointed out in Sec.~\ref{sec:symb-weight-dist}, we use Monte
Carlo simulations only to obtain the PAM4-symbol-error weight
distributions of the information bits at the output of the inner
decoders (i.e., the distribution of $U_j$ in
\eqref{eqn:symb-err-dist} and \eqref{eqn:symb-err-dist2}) for
various eBCH/SPC code parameters and input SNR values.
For each inner code parameter, we precomputed the
error-weight distributions at input SNRs between 12 and 18~dB
in increments of 0.05~dB.
The error-weight distribution at an intermediate input SNR
is approximated
by linearly interpolating between the error-weight distributions
corresponding to the two nearest multiples of 0.05~dB.
The error-weight distributions are stored in a
database, which is accessed as needed while using
\eqref{eqn:cond-symb-weight}--\eqref{eq:unionbound} to estimate
the required input SNR that achieves $10^{-13}$ post-FEC FER.

\subsection{Code Search}
\label{sec:code-search}

A total of about 87 million code combinations were evaluated, from
which we determine the Pareto-efficient code combinations for
each code rate listed in Sec.~\ref{sec:criteria}.  Due to the
large search space spanning many code rates, we only show and
discuss the code search results for codes of rate-0.88 in this
paper. A complete listing of
Pareto-efficient codes of rates 0.75, 0.76,\,\ldots, 0.93 is given
online~\cite{sukmadji-pareto}.

\begin{figure}[t]
\centering
\includegraphics{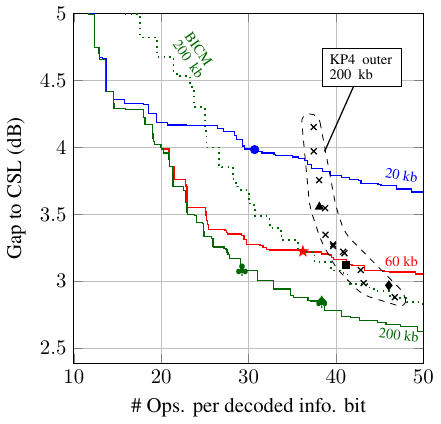}
\caption{Performance-complexity-latency trade-offs of
Pareto-efficient rate-0.88 ($\pm$ 0.005) concatenated RS-SDBCH
codes with KP4 and other RS outer codes.  Each curve shows
Pareto-efficient operating point with the maximum latency
indicated.  Also included are the Pareto-efficient operating
points for BICM with maximum latency of 200 kilobits.  The
parameters of some selected data points with markers other than
``$\times$'' are listed in Table~\ref{tab:rate-88}.}
\label{fig:pareto-88}
\end{figure}

The performance-complexity-latency trade-off curves for
Pareto-efficient codes of rate 0.88 ($\pm$ 0.005) are shown in
Fig.~\ref{fig:pareto-88} with the code parameters of some selected
data points listed in Table~\ref{tab:rate-88}.  Based on the
results, we conclude that KP4 is not the best choice of outer code
for our concatenated scheme, as there are many combinations of
outer RS and inner SDBCH codes that attain better performance at
the same complexity, or lower complexity at the same performance.

We notice that all Pareto-efficient rate-0.88 code parameters use
MLC instead of BICM.  Indeed, as shown in
Fig.~\ref{fig:pareto-88}, we observe gaps between the overall
Pareto-efficient curve (which comprises of only MLC) and that only
for BICM at maximum latency of 200 kilobits.
We believe that this is due to two reasons:
\begin{itemize}
\item Only half of the bits, i.e., the LSBs, are being decoded by
the inner code in the MLC scheme as opposed to both MSBs and LSBs
in BICM, therefore the number of required elementary inner
decoding operations is effectively halved.  Even if MLC requires
the MSBs to be conditionally demapped, the total number of
operations is still lower than that for BICM.
\item For a fixed inner rate, the LSBs in the MLC scheme are
protected by a code with roughly half the length of that used in
BICM, while the number of parity bits remains about the same.
Thus, the rate of the BCH code used to decode the LSBs in MLC is
lower than that for decoding the MSBs and LSBs in BICM, which
generally leads to better error-correcting performance for the
LSBs of the MLC scheme. Given a corrected LSB, conditional
demapping the corresponding MSB rarely results in an incorrect
bit.
\end{itemize}
We note that this observation that MLC outperforms BICM in a
concatenated coding scheme with higher-order modulation agrees
with previous work of, e.g.,
\cite{barakatain-higherordermodulation,mehmood,matsumine-jlt,matsumine-tcom}.

\begin{figure}[t]
\centering
\includegraphics{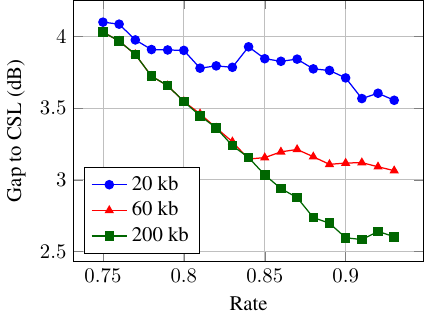}
\caption{Gap to the CSL of Pareto-efficient codes with varying
rates and complexity score of approximately 40.}
\label{fig:pareto-comp40}
\end{figure}

Fig~\ref{fig:pareto-comp40} shows the gap to the CSL of
Pareto-efficient codes over various code rates with complexity
score of approximately 40 and maximum latencies of 20, 60, and 200
kilobits.  As expected, the gap to CSL generally goes down as the
code rate increases. However, the curves are not monotonically
decreasing at higher rates. This is because higher-rate codes need
longer codes and the code search criteria specified in
Sec.~\ref{sec:criteria} limits the possible combinations of inner
and outer code parameters.  Thus, all possible combinations of
inner and outer code parameters for a particular code rate may not
necessarily include a code that outperforms the best code at a
lower rate.  This can be resolved by, for example, increasing the
maximum latency. Indeed, when the maximum latency is increased,
the possible combinations of inner and outer code parameters also
increase, which may give better code parameters than those at a
lower maximum latency.

Finally, the reader is reminded that the conclusions that we have
drawn in this paper are based on our
code search space as well as our choice of the figures of merit.
These conclusions may change
were we to enlarge the code search space and/or change the figures
of merit.

\begin{table}[t]
\centering
\caption{Parameters of Some Selected Pareto-Optimal Rate-0.88 Codes With KP4 and Other RS Outer Codes ($B=10$)}
\renewcommand{\tabcolsep}{3pt}
\begin{tabular}{|c|c|c|c|c|c|c|c|c|c|c|c|}\hline
$M$ & $N$ & $T$ & $m$ & $n$ & $b$ & $t$ & $J$ & Lat. & Compl. & Gap (dB) & Type\\\hline
{\color{blue}${}^\bullet$}4 & 458 & 10 & 229 & 47 & 6 & 1 & 5 & 19923 & 30.69 & 3.984 & MLC\\\hline
${}^\blacktriangle$10 & 544 & 15 & 544 & 57 & 6 & 1 & 2 & 58208 & 38.10 & 3.556 & MLC\\
{\color{red}${}^\bigstar$}8 & 689 & 14 & 689 & 47 & 6 & 1 & 5 & 59943 & 36.22 & 3.220 & MLC\\\hline
${}^\blacksquare$22 & 544 & 15 & 544 & 125 & 7 & 2 & 4 & 127840 & 41.14 & 3.121 & MLC\\
{\color{green!40!black}${}^\clubsuit$}21 & 546 & 7 & 546 & 127 & 7 & 3 & 5 & 126672 & 29.25 & 3.083 & MLC\\\hline
${}^\blacklozenge$25 & 544 & 15 & 544 & 142 & 8 & 2 & 6 & 145248 & 46.05 & 2.968 & MLC\\
{\color{green!40!black}${}^\spadesuit$}14 & 991 & 10 & 991 & 85 & 7 & 2 & 6 & 153605 & 38.38 & 2.845 & MLC\\\hline
\end{tabular}
\label{tab:rate-88}
\end{table}

\section{Conclusion}
\label{sec:conclusion}
In this paper, we have expanded our previous work
\cite{sukmadji2024} to derive a semi-analytical formula to predict
the post-FEC FER of a concatenated coding scheme with outer RS
codes and inner SDBCH codes with PAM4 modulation scheme under both
the BICM and MLC coded modulation architectures. We have used the
formula to bypass time-consuming Monte Carlo simulations to
predict the required input SNR so that $10^{-13}$ post-FEC FER is
achieved. We have provided performance-complexity-latency curves
for codes of rate 0.88, with more extensive code-search results
given in~\cite{sukmadji-pareto}.

Future work may include extending the measure of complexity to
include energy dissipation of the decoder circuit. We may also
consider determining the performance-complexity-latency trade-offs
of the concatenated scheme using an even higher-order modulation
scheme such as PAM6 or PAM8.

Another important question is to consider the effect of errors arising
in the chip-to-module (C2M) electrical interface.  Whereas the outer RS
encoder and decoder typically reside in the host chip, the inner BCH
encoder and decoder are connected to the host chip via the C2M serial
link, which itself will introduce bit errors (with a bit error rate on
the order of $10^{-5}$ \cite{lusted,li}) in addition to those arising
from transmission through the optical channel.  While this question is
outside the scope of the present paper, it would be interesting to see
how these additional errors affect the Pareto-efficient code
combinations.

\appendix
\subsection{Complexity of Chase Decoding of BCH codes}
\label{app:chase-complexity}
Recall from Sec.~\ref{sec:complexity} that \emph{elementary
decoding operations} constitute (real-valued and integer)
additions and multiplications, comparisons, symbol-wise XOR, and
table lookups.  There are a few steps to consider when decoding a
BCH($n,k,t$) code using Chase decoding with $J$ test bits. The
numbers of the elementary operations needed for some steps in
hard-decision BCH decoding are the same as that in
\cite[Tab.~III]{sukmadji2024}, namely:
\begin{itemize}
	\item initial syndrome computation: $nt$,
	\item key equation solver: $0,11,23$ for $t=1,2,3$, respectively,
	\item polynomial root finding: $0,10,37$ for $t=1,2,3$, respectively,
	% \item Bit correction (BC): $t$
\end{itemize}
Let $\mathcal{I}\subseteq[n]$ be the set of $J$ least reliable bit
indices. The additional operations needed for Chase decoding are
listed below.

\subsubsection{LLR and Hard-Decision Computation}% (LH)}
We assume that we have memory storing the values of LLRs of both
MSBs and LSBs given the channel output.  Retrieval of these values
takes $n$ memory accesses. Furthermore, an additional $n$
comparisons are required to obtain the initial hard-decision input
bits given the LLRs by checking the signs of the LLRs.

\subsubsection{Finding the Indices of the Least Reliable Bits}% (FJ)}
We assume a sorted linked list in decreasing reliability scores
with $J$ entries with the head pointing to the node corresponding
to the node with the largest of the $J$ running lowest absolute
values of the LLRs.  The worst-case complexity happens when the
reliabilities of the $n$ input bits are already sorted in
decreasing order in the first place.

For the worst-case complexity of the initialization of a
length-$J$ sorted linked list, it takes
$0+1+\cdots+(J-1)=\frac{J(J-1)}{2}$ operations for finding the
right position, plus $3J$ operations to create nodes and fix the
pointers between the previous and next nodes.  After the
initialization, for each of the remaining $n-J$ bits, since the
incoming LLRs are sorted in decreasing order, the linked list will
delete the node that the head pointer points to, and it will
perform up $J$ comparisons before creating the node at the tail of
the linked list. This will take at most $J(n-J)$ comparisons plus
$5(n-J)$ additional operations for creating new nodes, fixing the
pointers at the tail of the linked list, moving the head pointer,
and deleting the node to which the head pointer previously
pointed. In total, there are $(n-J)(J+5)+\frac{J(J-1)}{2}+3J$
operations.

\subsubsection{Test Pattern Generation}% (PG)}
We assume that the $2^J$ test patterns are generated such that the
Hamming distance between one test pattern with the next is one, as
in Gray coding.  The number of steps needed to generate those test
patterns is $2^J-1$.

\subsubsection{Test Syndrome Generation}% (SG)}
We assume that the syndrome corresponding to one-bit patterns in
$\mathcal{I}$ is readily available by reading off the
corresponding columns of the parity check matrix. We also assume
that the $bt$ bits of the syndromes comprise of $t$ $b$-bit
symbols over $\mathbb{F}_{2^b}$.  Since each test pattern is
generated in a Gray-coded fashion, generating the succeeding test
syndromes will then just take $t$ steps. Thus, the complexity
required for this step is $t\left(2^J-1\right)$.

\subsubsection{Analog Weight Computation}% (AW)}
For each successful decoding attempt, the decoder keeps track of
the position of the 1s in the test pattern as well as the
positions of bits to flip from the RF operation. Thus, suppose
that the test pattern has Hamming weight $s$. The worst case is
when the decoder declares $t$ errors in positions disjoint with
the positions of the ones in the test pattern.  The analog weight
will then be computed by adding $s+t$ absolute LLRs, which
requires $s+t-1$ operations.  Thus, the worst-case complexity for
computing the analog weight is
\[
\binom{J}{0}(t-1)+\binom{J}{1}t+\cdots+\binom{J}{J}(t+J-1)=\sum_{s=0}^J \binom{J}{s}(t+s-1).
\]

For extended Hamming inner code, i.e., $t=1$, only half of the
operations are done since the decoder will only attempt to decode
when the Hamming weight of the test input is odd. Thus, the
worst-case complexity for computing the analog weight of
Chase-decoded extended Hamming code is
\[
\max\left\{\sum_{s=0}^{\left\lfloor J/2\right\rfloor} \binom{J}{2s}(2s),\sum_{s=0}^{\left\lfloor J/2\right\rfloor} \binom{J}{2s+1}(2s+1)\right\}.
\]

\subsubsection{Finding the Lowest Analog Weight}% (FL)}
Up to $2^J-1$ comparisons comparisons are needed to find the lowest
analog weight, or $\frac{2^J}{2}-1$ in the case of extended
Hamming codes.

\subsubsection{Bit Correction}% (BC)}
The largest number of bits to be corrected in Chase decoding is
when the decoder select the output codeword that corresponds to
that with the test pattern that flips all $J$ test bits plus $t$
from the decoder. Thus, the worst-case complexity of bit
correction is $J+t$.

\subsubsection{Demapping the MSBs (For MLC)} %(DM)}
For MLC, the MSBs are obtained by conditionally demapping the
channel outputs based on
the corrected LSBs. This takes two operations per MSB: checking
whether the corresponding corrected LSB is zero or one, then demap
based on the appropriate set partition. Thus, it takes $2k$
operations in total.

\subsection{Decoding SPC Codes Using Wagner Decoding}
\label{app:wagner}

We assume that the first $n-1$ bits of an SPC code are information
bits and the last bit is the parity bit.  From the channel output
$\mathbf{y}$, the log-likelihood ratio (LLR) for
each of the $n$ bits
$\boldsymbol{\ell}=(\ell_{0},\ldots,\ell_{n-1})$ is computed
and then the hard-decision word
$\mathbf{x}=\left(x_0,\ldots,x_{n-1}\right)$ is obtained
by examining the signs of the LLRs.

Next, the overall parity $P$ of $\mathbf{x}$ is computed, where
$P=x_0\oplus\cdots\oplus x_{n-1}$. If $P=0$, then the decoder
returns the information bits of $\mathbf{x}$, i.e.,
$(x_0,\ldots,x_{n-2})$.  Otherwise, the decoder will find the
index of the least reliable bit, i.e.,
$q^*=\argmin_q\left\{|\ell_0|,\ldots,|\ell_{n-1}|\right\}$.  If
$q^*=n-1$, then the decoder returns $(x_0,\ldots,x_{n-2})$.
Otherwise, the decoder returns
\[
(x_0,\ldots,x_{q-1},x_q\oplus 1,x_{q+1},\ldots,x_{n-2}).
\]

\IEEEtriggeratref{22}


\begin{thebibliography}{10}
\providecommand{\url}[1]{#1}
\csname url@samestyle\endcsname
\providecommand{\newblock}{\relax}
\providecommand{\bibinfo}[2]{#2}
\providecommand{\BIBentrySTDinterwordspacing}{\spaceskip=0pt\relax}
\providecommand{\BIBentryALTinterwordstretchfactor}{4}
\providecommand{\BIBentryALTinterwordspacing}{\spaceskip=\fontdimen2\font plus
\BIBentryALTinterwordstretchfactor\fontdimen3\font minus
  \fontdimen4\font\relax}
\providecommand{\BIBforeignlanguage}[2]{{%
\expandafter\ifx\csname l@#1\endcsname\relax
\typeout{** WARNING: IEEEtran.bst: No hyphenation pattern has been}%
\typeout{** loaded for the language `#1'. Using the pattern for}%
\typeout{** the default language instead.}%
\else
\language=\csname l@#1\endcsname
\fi
#2}}
\providecommand{\BIBdecl}{\relax}
\BIBdecl

\bibitem{sukmadji-ofc25}
\BIBentryALTinterwordspacing
A.~Y. Sukmadji and F.~R. Kschischang, ``Performance-complexity-latency
  trade-offs of concatenated {RS-SDBCH} codes,'' accepted for
presentation at the 2025
  IEEE Opt.\ Fiber.\ Commun.\ Conf. (OFC), San~Francisco, CA, USA,
Mar.~30--Apr.~3, 2025. [Online]. Available:
  \url{https://arxiv.org/abs/2410.16535}
\BIBentrySTDinterwordspacing

\bibitem{chang}
F.~Chang and S.~Bhoja, ``New paradigm shift to {PAM4} signalling at 100/400{G}
  for cloud data centers: A performance review,'' in \emph{Eur. Conf. Opt.
  Commun. (ECOC)}, Gothenburg, Sweden, Sep.~17--21, 2017, pp. 1--3.

\bibitem{troncoso}
M.~Troncoso-Costas, D.~Dass, C.~Browning, F.~J. Diaz-Otero, C.~G.~H.
  Roeloffzen, and L.~P. Barry, ``Intra-data centre flexible {PAM} transmission
  system using an integrated {InP-Si${}_3$N${}_4$} dual laser module,''
  \emph{{IEEE} Photon. J.}, vol.~14, no.~1, pp. 1--6, Feb. 2022.

\bibitem{buchali}
F.~Buchali, X.-Q. Du, K.~Schuh, S.~T. Le, M.~Gr\"{o}zing, and M.~Berroth, ``A
  {SiGe} {HBT} {BiCMOS} 1-to-4 {ADC} frontend enabling low bandwidth
  digitization of 100 {G}baud {PAM}4 data,'' \emph{J. Lightw. Technol.},
  vol.~38, no.~1, pp. 150--158, Jan.~1, 2020.

\bibitem{vanveen}
D.~van Veen and V.~Houtsma, ``Real-time validation of downstream 50{G}/25{G}
  and 50{G}/100{G} flexible rate {PON} based on {M}iller encoding, {NRZ}, and
  {PAM4} modulation,'' \emph{{IEEE} J. Opt. Commun. Netw.}, vol.~15, no.~8, pp.
  C147--C154, Aug. 2023.

\bibitem{agrell2024}
E.~Agrell, M.~Karlsson, F.~Poletti, S.~Namiki, X.~Chen, L.~A. Rusch,
  B.~Puttnam, P.~Bayvel, L.~Schmalen, Z.~Tao, F.~R. Kschischang, A.~Alvarado,
  B.~Mukherjee, R.~Casellas, X.~Zhou, D.~van Veen, G.~Mohs, E.~Wong,
  A.~Mecozzi, M.-S. Alouini, E.~Diamanti, and M.~Uysal, ``Roadmap on optical
  communications,'' \emph{J. Optics}, vol.~26, no.~9, pp. 1--64, Jul. 2024.

\bibitem{ieee8023df}
``{IEEE} standard for {E}thernet amendment 9: Media access control parameters
  for 800 {G}b/s and physical layers and management parameters for 400 {G}b/s
  and 800 {G}b/s operation,'' IEEE, Standard 802.3df, Feb. 2024.

\bibitem{nagarajan}
R.~Nagarajan, A.~Martino, D.~A. Morero, L.~Patra, C.~Lutkemeyer, and M.~A.
  Castrillón, ``Recent advances in low-power digital signal processing
  technologies for data center applications,'' \emph{J. Lightw. Technol.},
  vol.~42, no.~12, pp. 4222--4232, Jun.~15, 2024.

\bibitem{willner}
S.~Kumar, G.~Papen, K.~Schmidtke, and C.~Xie, ``Intra-data center
  interconnects, networking, and architectures,'' Ch.~14 in \emph{Optical Fiber
  Telecommunications VII}, A.~E. Willner, Ed.\hskip 1em plus 0.5em minus
  0.4em\relax Academic Press, 2020, pp. 627--672.

\bibitem{szczerba}
K.~Szczerba, P.~Westbergh, J.~Karout, J.~S. Gustavsson, A.~Haglund,
  M.~Karlsson, P.~A. Andrekson, E.~Agrell, and A.~Larsson, ``4-{PAM} for
  high-speed short-range optical communications,'' \emph{{IEEE} J. Opt. Commun.
  Netw.}, vol.~4, no.~11, pp. 885--894, Nov. 2012.

\bibitem{zehavi}
E.~Zehavi, ``8-{PSK} trellis codes for a {R}ayleigh channel,'' \emph{{IEEE}
  Trans. Commun.}, vol.~40, no.~5, pp. 873--884, May 1992.

\bibitem{caire}
G.~Caire, G.~Taricco, and E.~Biglieri, ``Bit-interleaved coded modulation,''
  \emph{{IEEE} Trans. Inf. Theory}, vol.~44, no.~3, pp. 927--946, May 1998.

\bibitem{wachsmann}
U.~Wachsmann, R.~Fischer, and J.~Huber, ``Multilevel codes: theoretical
  concepts and practical design rules,'' \emph{{IEEE} Trans. Inf. Theory},
  vol.~45, no.~5, pp. 1361--1391, Jul. 1999.

\bibitem{ungerboeck}
G.~Ungerboeck, ``Channel coding with multilevel/phase signals,'' \emph{{IEEE}
  Trans. Inf. Theory}, vol.~28, no.~1, pp. 55--67, Jan. 1982.

\bibitem{imai}
H.~Imai and S.~Hirakawa, ``A new multilevel coding method using
  error-correcting codes,'' \emph{{IEEE} Trans. Inf. Theory}, vol.~23, no.~3,
  pp. 371--377, May 1977.

\bibitem{smith}
B.~P. Smith, A.~Farhood, A.~Hunt, F.~R. Kschischang, and J.~Lodge, ``Staircase
  codes: {FEC} for 100 {G}b/s {OTN},'' \emph{J. Lightw. Technol.}, vol.~30,
  no.~1, pp. 110--117, Jan.~1, 2012.

\bibitem{barakatain-higherordermodulation}
M.~Barakatain, D.~Lentner, G.~B\"{o}cherer, and F.~R. Kschischang,
  ``Performance-complexity tradeoffs of concatenated {FEC} for higher-order
  modulation,'' \emph{J. Lightw. Technol.}, vol.~38, no.~11, pp. 2944--2953,
  Jun.~1, 2020.

\bibitem{mehmood}
T.~Mehmood, M.~P. Yankov, S.~Iqbal, and S.~Forchhammer, ``Flexible multilevel
  coding with concatenated polar-staircase codes for {M}-{QAM},'' \emph{{IEEE}
  Trans. Commun.}, vol.~69, no.~2, pp. 728--739, Feb. 2021.

\bibitem{yoshida}
T.~Yoshida, I.~Kudo, K.~Ishii, H.~Yoshida, H.~Shimizu, S.~Hirano, Y.~Konishi,
  M.~Karlsson, and E.~Agrell, ``High-speed multilevel coded modulation and soft
  performance monitoring in optical communications,'' in \emph{Opt. Fiber
  Commun. Conf. Exhibit. (OFC)}, San Diego, CA, USA, Mar.~24--28 2024, pp. 1--3.

\bibitem{sukmadji2024}
A.~Y. Sukmadji and F.~R. Kschischang, ``Performance-complexity-latency
  trade-offs of concatenated {RS-BCH} codes,'' \emph{{IEEE} Trans. Commun.},
  vol.~72, pp. 3829--3841, Jul. 2024.

\bibitem{chase}
D.~Chase, ``Class of algorithms for decoding block codes with channel
  measurement information,'' \emph{{IEEE} Trans. Inf. Theory}, vol.~18, no.~1,
  pp. 170--182, Jan. 1972.

\bibitem{lentner23}
D.~Lentner, E.~B. Yacoub, S.~Calabr\`{o}, G.~B\"{o}cherer, N.~Stojanovi\'{c},
  and G.~Kramer, ``Concatenated forward error correction with {KP4} and single
  parity check codes,'' \emph{J. Lightw. Technol.}, vol.~41, no.~17, pp.
  5641--5652, Sep.~1, 2023.

\bibitem{yang}
L.~Yang, J.~Tian, B.~Wu, Z.~Wang, and H.~Ren, ``An {RS-BCH} concatenated {FEC}
  code for beyond 400 {G}b/s networking,'' in \emph{IEEE Comput. Soc. Annu.
  Symp. VLSI (ISVLSI)}, Nicosia, Cyprus, Jul.~4--6, 2022, pp. 212--216.

\bibitem{wang2023}
X.~Wang, X.~He, and H.~Ren, ``Advanced {FEC} for 200 {G}b/s transceiver in 800
  {GbE} and 1.6 {TbE} standard,'' \emph{{IEEE} Commun. Standards Mag.}, vol.~7,
  no.~3, pp. 56--62, Sep. 2023.

\bibitem{ieee8023}
``802.3-2022 -- {IEEE} standard for {E}thernet,'' IEEE Standard, pp. 1--7025,
  Jul. 2022.

\bibitem{bhoja}
\BIBentryALTinterwordspacing
S.~Bhoja, V.~Parthasarathy, and S.~Wang, ``{FEC} codes for 400 {G}bps
  802.3bs,'' IEEE P802.3bs 200 GbE \& 400 GbE Task Force, Nov. 2014. [Online].
  Available:
  \url{https://www.ieee802.org/3/bs/public/14_11/parthasarathy_3bs_01a_1114.pdf}
\BIBentrySTDinterwordspacing

\bibitem{chagnon}
M.~Chagnon, S.~Lessard, and D.~V. Plant, ``336 {G}b/s in direct detection below
  {KP}4 {FEC} threshold for intra data center applications,'' \emph{{IEEE}
  Photon. Technol. Lett.}, vol.~28, no.~20, pp. 2233--2236, Oct.~15, 2016.

\bibitem{sukmadji-pareto}
\BIBentryALTinterwordspacing
A.~Y. Sukmadji and F.~R. Kschischang, ``List of {P}areto-efficient concatenated
  {RS-SDBCH} codes,'' 2024. [Online]. Available:
  \url{https://www.comm.utoronto.ca/~asukmadji/rs_sdbch_pareto.html}
\BIBentrySTDinterwordspacing

\bibitem{silverman}
R.~A. Silverman and M.~Balser, ``Coding for constant-data-rate systems-{P}art
  {I}. a new error-correcting code,'' \emph{Proc. IRE}, vol.~42, no.~9, pp.
  1428--1435, Sep. 1954.

\bibitem{qu2024}
\BIBentryALTinterwordspacing
Y.~Qu, A.~Tasbihi, and F.~R. Kschischang, ``Constituent automorphism decoding
  of {R}eed--{M}uller codes,'' Sep. 2024. [Online]. Available:
  \url{https://arxiv.org/abs/2409.03700}
\BIBentrySTDinterwordspacing

\bibitem{matsumine-jlt}
T.~Matsumine, M.~P. Yankov, and S.~Forchhammer, ``Geometric constellation
  shaping for concatenated two-level multi-level codes,'' \emph{J. Lightw.
  Technol.}, vol.~40, no.~16, pp. 5557--5566, Aug.~15, 2022.

\bibitem{matsumine-tcom}
T.~Matsumine, M.~P. Yankov, T.~Mehmood, and S.~Forchhammer, ``Rate-adaptive
  concatenated multi-level coding with novel probabilistic amplitude shaping,''
  \emph{{IEEE} Trans. Commun.}, vol.~70, no.~5, pp. 2977--2991, May 2022.

\bibitem{lusted}
\BIBentryALTinterwordspacing
K.~Lusted and M.~Nowell, ``{BER} targets for type 1 and
type 2 {PHY}s,'' IEEE P802.3dj Task Force, May. 2023. [Online].
  Available:
  \url{https://www.ieee802.org/3/dj/public/23_05/lusted_3dj_03a_2305.pdf}
\BIBentrySTDinterwordspacing

\bibitem{li}
\BIBentryALTinterwordspacing
M.~P.~Li, J.~Lim, A.~Ghiasi, K.~Gopalakrishnan, I.~Lyubomirsky, P.~Dawe, E.~Frlan,
and M.~Kimber, ``Baseline proposal for `100 {G}b/s, 200 {G}b/s, and 400 {G}b/s chip-to-module
attachment unit interface
(100{GAUI}-1, 200{GAUI}-2, and 400{GAUI}-4)','' IEEE P802.3ck Task Force, Mar. 2019. [Online].
  Available:
  \url{https://www.ieee802.org/3/ck/public/19_03/li_3ck_02b_0319.pdf}
\BIBentrySTDinterwordspacing
\end{thebibliography}
\end{document}